# Hacking the Brain:

## Triggering Neuroplasticity for Enhancing Musical Talent: A study on Monkey and Human behavior after Exposure to Videogames and Visual/Auditory Stimuli to Increase Musical Abilities through Neuroplasticity

Lucas Agudiez Roitman

Prof. Poppy Crum

Stanford's Center for Computer Research in Music and Acoustics

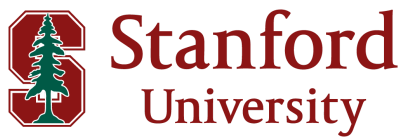

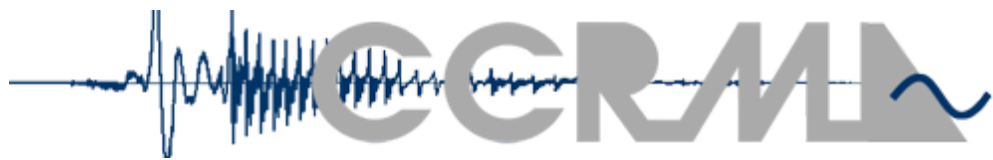

**Abstract**

In this paper, we analyze the cognitive improvements that can be achieved through hacking the brain through the use of multiple methods to enhance neuroplasticity.

Exposure to gaming, for example, has proven conducive for learning real-world abilities through auditory and visual stimuli. We will discuss cortical magnification and receptive field sizes, as well as topographic brain maps in the context of neuroplasticity. We finally propose more studies to be performed to improve musical talent and musical abilities.

**Cognitive Improvements**

Some cognitive improvements have been achieved through exposure to gaming.[1] They also correlate with motor changes. Games influence perceptual processing, as demonstrated in a number of studies[2]; cognitive improvements influence central and peripheral attention skills in addition to multiple object tracking spatial resolution.[3] These cognitive improvements are correlated with visual acuity, task switching, and object tracking as well as other types of motor changes.[4]

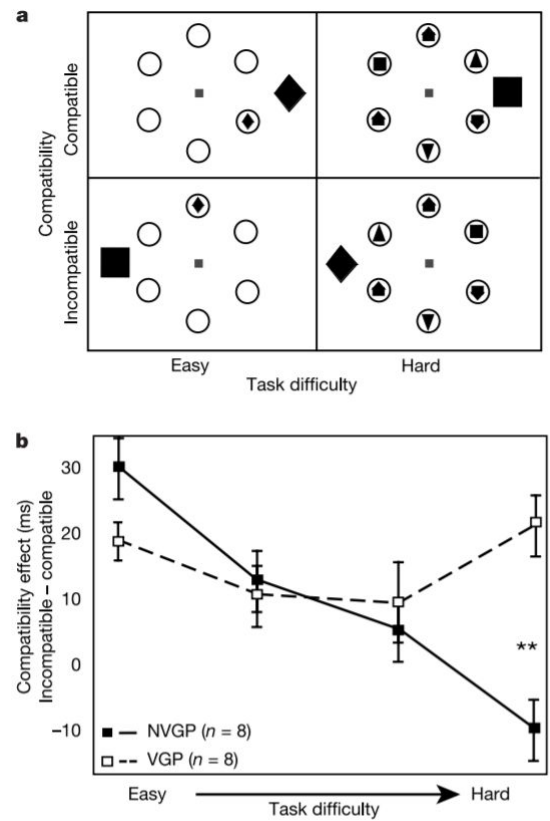

Not all these improvements can be generalized to other tasks. There are some domains that specific changes cannot be generalized to. Only some specific skills can be generalized to other tasks, such as perceptual skills needed by professional plane pilots. One study discovered that experienced video gamers were better at spatial navigation than non-experienced members only in the game; the study found no difference in ability level between the two groups.[5]

---

[1] Bavelier, Daphne, et al. "Brains on video games." Nature reviews neuroscience 12.12 (2011): 763.
[2] Green, C. Shawn, and Daphne Bavelier. "Action video game modifies visual selective attention." Nature 423.6939 (2003): 534.
[3] Green, C. Shawn, and Daphne Bavelier. "Action-video-game experience alters the spatial resolution of vision." Psychological science 18.1 (2007): 88-94.
[4] Bavelier, Daphne, and Helen J. Neville. "Cross-modal plasticity: where and how?." Nature Reviews Neuroscience 3.6 (2002): 443.
[5] Bavelier, Daphne, et al. "Brains on video games." Nature reviews neuroscience 12.12 (2011): 763.

## Helpful and harmful cases

There are some cases where gaming exposure is helpful and others where it is harmful for the subject. Game exposure examples may not all have positive, or negative effects on the gamer. Fast, concentration-demanding video games, form new neural pathways in response to the game. Gamers can begin developing rapid response times as they more quickly filter out visual distractors. A negative effect of these fast, intense games is that gamers are susceptible to developing ADHD as peripheral information more easily catches their attention, creating distraction.

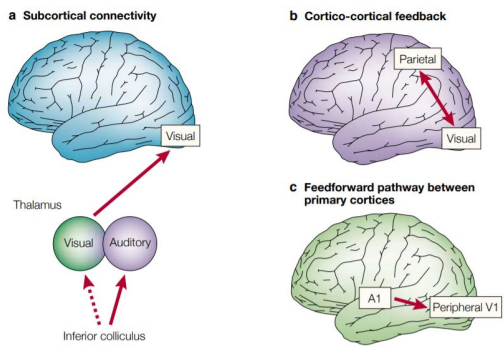

Additionally, social games teach the player to behave more collaboratively, or helpfully in these environments, whereas a violent game may teach gamers more violent behavior. Children have been identified as the group the more impressionable to the behavior in these games.

A danger of taking up gaming is that it can become addictive, and begin taking up more time in the day than anticipated. Addiction is a distractor and can cause gamers to shirk responsibility. This addiction can intensify into other anti-social behavioral expressions like depression, and anxiety.[6]

Skills involving attention are a part of the solution and the problem discussed earlier. Improving attention to periphery and developing a wider field of view result from playing action games. These skills improve performance in the game setting, but can be a disadvantage in the learning environment because of this meticulous attention to detail.

## Differences in new generations

What is different about the young gaming generation from past generations is that they have grown up with entertainment and stimulation delivered by the nearest monitor. This monitor or screen delivered media have captured their attention and made them hungry for more. The younger generation's constant digestion of this digital media has changed how their brains engage in life when compared to older generations.

---

[6] Bavelier, Daphne, et al. "Brains on video games." Nature reviews neuroscience 12.12 (2011): 763.

The rise in digital media exposure offers an increase of possibilities in treating certain psychiatric or neurological illnesses. This paper by Bavelier et all asserts that medical claims about training programs or learning tools should undergo a formal review process. Yet, this begs the question of whether it is feasible to generalize to all types of screen-delivered information from content, games, and media.

## What games provide that is conducive to learning

Video games motivate individuals through a sort of benefit-reward based way of learning that drives a positive, affirming learning model. Brain training exercises are conducive because they can be corrective but create a more positive experience for the user.

They encourage the learner through enthusiastic reassurance and guidance. Learning through gaming is a process that is active in nature, as it gives you a play-by-play of reinforcement. Over time reinforcement of the right skills allows the player to master the material. Games allow the user to practice what they learn in different environments.

## Impulsivity for children in gaming

In thinking about gaming and its effect on children, we may be tempted to ban our children from it because of the perceived dangers like addiction, ADHD, or violent behavior.[7] However, games create enormous opportunity because of their enthusiastic way of teaching material and testing the material in different contexts in an active, engaging manner. Game development and game design can be tailored to maximize the positive results of gaming like attention to detail and faster processing, and perceived dangers like violence can be avoided by only allowing games meant to be played by children of a certain age. Game usage monitoring can be used to prevent

| Instrument/task | Measure | HV | VG | T | P-value |
| --- | --- | --- | --- | --- | --- |
| UPPS | | 128.37 (20.93) | 136.79 (29.60) | 1.56 | 0.12 |
| BDI | | 4.29 (5.03) | 7.61 (5.28) | 2.02 | 0.05 |
| DDT | K-value | 0.02 (0.04) | 0.07 (0.1) | 2.86 | 0.006 |
| IST | Boxes opened | 17.89 (5.72) | 14.13 (5.74) | 2.07 | 0.04 |
| Premature responding | Premature responding | 6.82 (5.29) | 10.36 (9.12) | 1.71 | 0.09 |
| | Motivational Index | 0.13 (0.16) | 0.18 (0.21) | −0.97 | 0.34 |
| Response inhibition | GoRT | 367.22 (81.04) | 366.27 (100.23) | 0.04 | 0.97 |
| | SSRT | 160.81 (49.95) | 154.27 (31.49) | 0.52 | 0.60 |

Abbreviations: HV = healthy volunteers; VG = pathological gamers; UPPS = UPPS Impulsive Behaviour Scale; BDI = Beck Depression Inventory-II; DDT = Delay Discounting Task; IST = Information Sampling Task.
doi:10.1371/journal.pone.0075914.t002

users from spending too much time gaming or from playing games that add little educational value.

## Cortical Magnification

Cortical magnification is the amount of neurons that are used to process stimuli. In vision, for example, if we see an object placed right in front of us, it will be captured by the center of the eye (the fovea) and will thus use about 50 cones per 100 micrometers, while if we see an object placed in the borders of our visual field, its image will be projected onto the perifovea and will thus only use 12 cones per 100 micrometers, resulting in a lower resolution.

Cortical magnification changes throughout the duration of an individual's life because of neuroplasticity. Learning and training your brain to perform different tasks many times leads to changes in cortical magnification such as in the case of monkeys that have improved their digital dexterity, and thus have modified their cortical magnification.

Two people listening to the same music might have very different experiences, such as if the music serves a different role in each of their lives. For example, as a violinist, listening to a song would be different than for a guitarrist, pianist, etc.

A violinist, for example, would pay more attention to the melody of the song and the pitch changes (and absolute pitches), while the pianist would probably be able to perceive the harmony and chords as more important and in more detail.

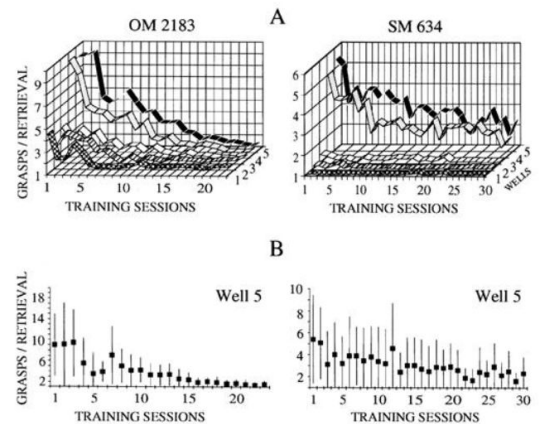

A DJ could be especially stimulated by the equalization and the percussionist by the beat. This is because the musicians' cortical magnification is different in each part of the brain, for each different task they usually perform, and thus that will lead to increased/decreased sensitivity for different stimuli.

In the paper "Assessment of sensorimotor cortical representation asymmetries and motor skills in violin players", violin players showed an increased right-left asymmetry of the somatosensory cortex, due to being trained in playing the violin.[8] In the paper "Representational plasticity in cortical area 3b paralleling tactual-motor skill acquisition in adult monkeys", it is shown that when a monkey is trained in digital dexterity, its cortical magnification will change and thus it will be more sensitive to the sensory inputs related to the task.[9]

---

[8] Schwenkreis, Peter, et al. "Assessment of sensorimotor cortical representation asymmetries and motor skills in violin players." European Journal of Neuroscience 26.11 (2007): 3291-3302.

## Receptive field sizes

Cortical magnification means that when there are many neurons used to interpret the stimuli, their receptive fields will be smaller than when there are fewer neurons. This is because fewer neurons will need to cover a bigger area of the visual image, for example, and will thus require bigger receptive fields to cover than increased area.

If cortical magnification increases the amount of neurons used for a specific task, it might also decrease the amount used for different ones. The cortical homunculus for a wind player would probably become bigger for the areas of the mouth (muscles and sensory nerves) used for the embouchure, whereas the cortical homunculus for the left-handed surgeon will likely have a bigger section for the left hand and specifically the muscles and sensors used for the profession.

## Topographic brain maps - cell clusters

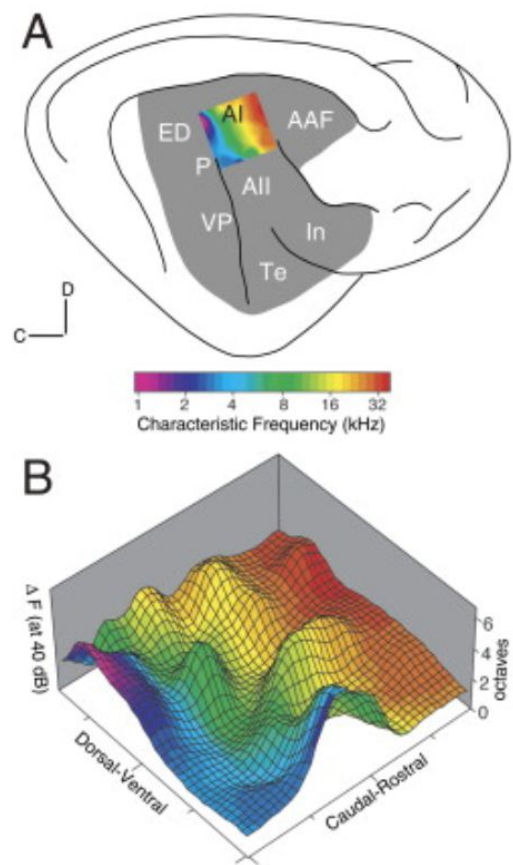

The benefits to having topographic maps in the brain are that in the case of learning new behaviour or adapting to new stimuli, the brain can periodically adapt the neuronal connections in the same area, making use of cells that have similar behaviors.

In "Auditory Cortex Mapmaking: Principles, Projections, and Plasticity" by Christoph E. Schreiner, and Jeffery A. Winer2, the authors explain how auditory topological maps can be useful for "plasticity" and adaptations.[10]

In violin players, topographic maps result in an enlargement of the left hand representation in the sensorimotor cortex of violin players, also providing evidence of neuroplasticity occurring in nearby areas of the cortex instead of using other areas of the brain, as researched by Peter Schwenkreis, Susan El Tom, Patrick Ragert, Burkhard Pleger, Martin Tegenthoff and Hubert R.

---

[9] Xerri, Christian, et al. "Representational plasticity in cortical area 3b paralleling tactual-motor skill acquisition in adult monkeys." Cerebral cortex 9.3 (1999): 264-276.
[10] Schreiner, Christoph E., and Jeffery A. Winer. "Auditory cortex mapmaking: principles, projections, and plasticity." Neuron 56.2 (2007): 356-365.

Dinse in the text "Assessment of sensorimotor cortical representation asymmetries and motor skills in violin players".

## Masking or distracting noise cancelling

The environment has increasing levels of noise. It is not important to perform cognitive tasks to understand sound while noise is masking some of the auditory components. One skill that could be useful to train would be identifying human voice and their words in a noisy environment. This could be useful for communicating in urban areas and might also be useful when trying to learn new languages.

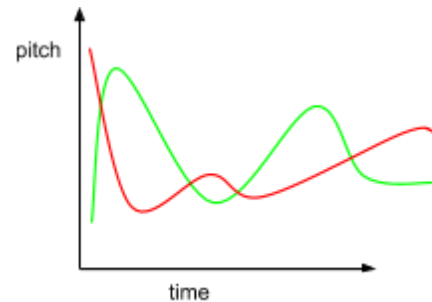

Another skill that one could train is identifying a specific instrument's melody that could be masked by other instruments in a song. This can allow a musician to better understand the roles of each instrument, and could extrapolate to the musician being able to identify its own instrument's sound in the song and allow to correct more easily.

We can train the brain by playing different instruments with different melodies, first on their own and then together in the first levels. In the more advanced ones, they would play together at the same time. And the user would have to input the melody by hand at the end, for a specific instrument.

At first, the game would test for the user to input a melody for a specific instrument. And at the end, it would repeat the same test, or group of tests, to measure the learning or neuroplasticity achieved. For both the game and diagnostic tests, we would use a violin and a piano playing different melodies. It could be synthesized or recorded live. The physiological changes would probably be cortical magnification, increasing the amount of neurons used to interpret the timbre of the different instruments.

## Conclusion

As we have seen in this paper, cortical magnification results in the increase in size of receptive fields, but also in the decrease in other areas in the brain. Therefore, it is important to proceed carefully so as to improve skills and talents that are useful without taking away from existing skills. Finally, we presented a method to hack the brain to improve a subject's musical talents and proposed a study for measuring this impact in brain neuroplasticity.